# Impact of Business Analytics and Decision Support Systems on e-commerce in SMEs

Ziad Almtiri[1,3], Shah J. Miah[2], and Nasimul Noman[3]

[1]Department of Management Information Systems, Taif University, Taif, KSA

[2] Newcastle Business School, University of Newcastle, NSW, Australia

[3] School of Information and Physical Sciences, University of Newcastle, NSW, Australia

ziad_almtiri@outlook.com, shah.miah@newcastle.edu.au,
nasimul.noman@newcastle .edu.au

*ABSTRACT* - With the advancement in the marketing channel, the use of e-commerce has increased tremendously therefore the basic objective of this study is to analyze the impact of business analytics and decision support systems on e-commerce in small and medium enterprises. Small and medium enterprises are becoming a priority for economies as by implementing some policies and regulations these businesses could encourage gain development on an international level. The objective of this study is to analyze the impact of business analytics and decision support systems on e-commerce in small and medium enterprises that investigate the relationship between business analytics and decision support systems in e-commerce businesses. To evaluate the impact of both on e-commerce the, descriptive analysis approach is adopted that reviews the research of different scholars who adopted different plans and strategies to predict the relationship between e-commerce and business analytics. The study contributes to the literature by examining the impact of business analytics in SMEs and provides a comprehensive understanding of its relationship with the decision support system. After analyzing the impact of business analytics and decision support system in SMEs, the research also highlights some limitations and provide future recommendations that are helpful to overcome these limitations.

*Keywords:* Business Analytics, Decision support system (DSS), E-commerce, ICT, SMEs, Enterprise resource planning (ERP)

## 1. INTRODUCTION

Electronic commerce offers electronic mechanism to improve the efficiency and effectiveness of different aspects of a product or service. It includes various digital aspects such as for design, production, marketing, and sales by utilizing current and emerging information technologies. In the contemporary era, small and medium enterprises (SMEs) must understand the adoption of information and communication technology (ICT) which is significant to improve their ways of development. Different E-Commerce channels are providing services to e-retailers who could select the products and add them to carts then check out based on their needs [1]. Furthermore, E-Commerce provides advanced methods that could help businesses to start economic growth and achieve the goals that are set out by the management. With the adoption of advanced communication technology, businesses could enable faster and cheaper communication and increase business operations that play a significant role in the development of e-businesses.

Business analytics is an extensive form of using data, statistical and quantitative analysis, and explanatory and predictive models that are used to make decisions and take actions in the business. Several businesses are striving to release their business policies and strategies for adopting better business analytics techniques that could make effective marketing campaigns, revenue forecasting, and automation system that will improve the performance of the business [2].

The other aspect is Decision Support System DSS which is an umbrella term used to explain the computer application that is used to improve the user's ability to decide for business [3]. The term is used to explain the system that is based on computer design and help the management to make a decision based on data knowledge and communication that could identify the problem and make the decision to solve these problems. DSS could be divided into different categories such as driven DSS, data-driven DSS, document driver DSS, knowledge driving DSS, model drive DSS, web-based DSS, and spreadsheet-based DSS.  Decision support systems are widely used in several organizations as small and medium enterprises have a common tendency to implement the experience and techniques that could gain from large organizations [4]. The early development of DSS was introduced in the 1970s and primarily generated management information from the operating system. The decision support system is commonly based on four components the model-based, the database, the user interface, and the user central [5].

The advancement in technology is used in the computation part of the data of an organization that provides easy access to data and flexible control that provides a variety of analytical tools. The basic objective of DSS is to improve the ability to decide by increasing the efficiency of management with reliable data [6]. The practices based on DSS could be hardly separated from the computer-based system as it is mostly integrated with the system based on operational databases, spreadsheets, report generators, and an executive support system. It includes the data and model-oriented system reporting system and other group decision support systems that are continuously growing the businesses of small and medium enterprises. These tools are used to solve the problem of SMEs just like large corporations as the use of information systems is helpful to overcome the problems and provide support to small and medium businesses according to the perspective of social and economic development. A few types of research have been conducted to analyze the decision support system requirement in the context of using information technology in SMEs. The management of SMEs is mostly disappointed with the use of software as there is the inability of adapting the software according to the needs of the customers therefore it may differentiate the problems of small businesses from large enterprises. SMEs usually have limited resources and a lack of skillful managerial staff therefore the failure risk is increased as compared to large corporations.

The implication of a decision support system in SMEs provides a productive domain and introduces an improved level of computer-based decision support. The management of SMEs is usually involved in day-to-day operations therefore it is required to include an analytical approach that will provide a strong decision based on business analytics and provide information to adopt the software that is suitable and could be effectively used by management to resolve the concerned problems.

E-commerce businesses required basic support for e-business transactions therefore E-commerce systems are required to become standard between suppliers and manufacturers. The E-Commerce system is based on computers and the internet, therefore, statistics represent that increasing interest in international shopping is also promoting the concept of e-commerce, therefore, manufacturers and retailers are required to improve their efforts to develop their operations in the international market [7]. The effects of manufacturers and retailers could improve the performance of SMEs as an E-commerce system permits them to perform the functions quickly at a low financial cost. The basic element in improving the efficiency of e-commerce is improved internal and external procedures as all the management activities and decisions are made according to the requirements of targeted customers therefore businesses could obtain optimal solutions for the problems and implement the procedural structure based on modeling and simulation. For organizations that are operating according to the E-commerce business model, it is considered appropriate to follow the business modeling as a central part of the projects.

The significance of e-commerce could be analyzed with the help of using a new business model that explains the benefits of the e-commerce system according to the perspective of SMEs [8]. The expanded scope and use of the new model highlight the needs of the customers and also provide information about the increase in demand that required deciding by the management. Information communication technology (ICT) is becoming a basic decision-making support system in SMEs as these technologies are based on an E-Commerce system and improved the ERP and CRM system of the business which is helpful to obtain the objectives of management. Currently, most organizations are based on the decision-making support modules and tools that are based on DSS and provide information that how these systems could be used in the development that is appropriate in specific conditions and could be classified as an intelligent decision support system.

### 1.1 Problem Statement

With the advancement in the internet, E-Commerce is considered a new way of doing business therefore it has the potential to change the economic activity and social environment of business. It has a huge impact on the performance of large corporations including their communication, finance, and retail trading, and holds a promise in specific areas to improve and put positive impacts. Similarly, small and medium organizations are also required to develop customized products and services that are based on the requirements of the customers, and the management of these companies is required to use proper market research and make a decision about the manufacturing or trading of the products and services that are according to the needs of the businesses. The paper emphasizes analyzing the problem related to finding the relationship between E-Commerce and decision support systems and finding the impact of business analytics and decision support systems on e-commerce networks adopted by small and medium enterprises.

### 1.2 Objectives of Study

By considering these facts the basic objective of the study is presented as analyzing the impact of business analytics and decision support systems on e-commerce at SME businesses. E-Commerce is the basic support of online shopping as it increases transactional efficiency and effectiveness and in the advanced era, it provides an understanding of the adoption of ICT and digital channels that are used by businesses. An E-commerce system is developed to support the management of SMEs. The result is required to review the literature to find out the following objectives:

- To learn about eCommerce channels adopted by small and medium enterprises.
- To find out the relationship between e-commerce business analytics and decision support systems and find out the impact of business analytics and decision support systems on e-commerce in SMEs.

The success and continuous development of small and medium organizations are significant for the growth of the economy but the problems could not be ignored therefore it is significant for the management of the organization to collect data and analyze the area where the problem may occur [8]. The use of information system could increase the attention of customers toward SMEs and also provides support to small businesses to prepare policies according to the analyzed data. The management of businesses is considered to use software packages that are helpful to improve the ability to meet the requirements of customers and reduce the failure risk that is appropriate for the development of SMEs. The improving use of e-commerce and marketing channels provides a great influence on the decisions of management. The influence of DSS could not be ignored on E-Commerce as it is a business activity that is developed to support business activities therefore the customers are required to meet their needs and requirements and it could not be possible without the implication of business. Data analytics provide information about the market data and figures that are analyzed with the help of a decision support system and interpret the results that help the management to make decisions [4].

While implementing The E-Commerce system, it is significant to link the system of the organization with the decision support system and other information systems that are used by the organization as all the changes in the e-commerce system could be made and it could make sure that the data is consistently following the instruction of management. Enterprise resource planning (ERP) is an appropriate strategy for international operations as it is based on different facilities and integrates the business procedures in different departments [2]. The major benefits of ERP are based on improving the coordination in the functional department and increasing efficiencies of doing business. The objective of this software is to collect the information that is processed by the management through a decision support system and make a decision that is beneficial for the organization. Briefly discuss the impact of decision support systems and business analytics, it would be explained that it provides a positive impact as it is used to improve the alignment of strategies and operations that improve productivity and reduce cost. The development of e-commerce in small and medium organizations could provide better relations with existing customers and also increase sales through reliable channels. These impacts are also helpful to develop the E-Commerce business on an international level and meeting the management standards of a large corporation that are based on achieving the goals of creating a relationship with customers, effective supply chain management, and operation management that could be refined with the help of information system management [1].

## 2. LITERTURE REVIEW

The impact of Business Analysis and Decision Support systems in e-commerce in SMEs is more critical with respect to company growth and development. A computerized information system that assists a business or business with its management, operations, and planning by organizing, collecting, and analyzing data is known as a decision support system (DSS)[1]. The majority of the time, DSSs are tasked with gathering data pertaining to sales, anticipated earnings, and stock levels. This information is utilized to create databases with pre-set links so sales data from different periods can be compared and analyzed[2].

A successful DSS makes it simple to incorporate data from a wide variety of sources, including, but not limited to, written records, raw data, management reports, business models, and the individual experiences of workers, and makes small enterprises more effective toward growth and development[3].

The DSS applications for Business Analysis and Decision Support systems at e-commerce in SMEs can be used in various contexts, including evaluating bids for engineering, agricultural, or rail construction projects, making medical diagnoses, checking credit loan applications, and the business of businesses[4].

Managers of Business Analysis and Decision Support systems at e-commerce in SMEs have various options for putting DSS programs to use. Most of the time, business strategists tailor a DSS to their requirements and then utilize it to analyze various processes and procedures. Sales and stock are examples of these processes, and both can benefit from the assistance that DSS programs can provide in improving the supply chain flow. Managers also benefit from DSS software since it assists them in planning and thinking about the potential implications of changes.

DSS may be beneficial because it can analyze a SMEs stock and assist the company in making the most of its asset listing[5].

In Business Analysis and Decision Support Systems at e-commerce in SMEs, Tracking trends and doing in-depth analyses of sales data to develop forecasts is another possible application of a decision support system. Because the system is simple to operate, planners have access to a diverse set of tools that can assist them in making judgments on customer data[6].

DSS software can also be used to make predictions about an SME's future or to obtain a comprehensive grasp of the factors that contribute to the expansion of a business. These two shows demonstrate the adaptability of the powerful program being used. This could be useful in challenging circumstances, the resolution of which needs a high level of financial prediction, such as when determining predicted and actual expenses and revenue[3].

Business Analysis and Decision Support Systems at e-commerce in SMEs or their Managers are continually working toward helping their companies expand. They achieve this goal by locating novel possibilities and resolving existing issues. Managers must make strategic decisions since managing a business in its various guises requires a variety of specific activities. When there is a lot of competition in a business sector, it can be difficult for managers to decide what actions to take. Regarding business-to-consumer (B2C) electronic commerce (e-commerce), managers may have difficulty deciding what actions to take when millions of people buy things directly from company websites. Therefore, DSS is the best approach to incorporate into their SMEs[7].

Studies showed that customers would look at several e-commerce websites before purchasing online to gain more knowledge and make educated decisions. When retailers and end users do business with each other directly over the internet, this type of transaction is known as "business-to-consumer" or "B2C" e-commerce. This is a pretty popular method utilized when conducting business activities online[8].

Therefore, if you want to meet the demands of any shift in the external business environment, you must create a robust business strategy. Taking care of these things in business is a popular choice for many businesses because it helps them maintain their current success level[9]. However, much small business (SME) owners and managers are poor planners whose short-term goals are unspecific and based on human intuition[10]. Such challenges can only be overcome with technological support and solutions, often out of reach for small businesses (SMEs). Researchers have also found that when organizations of all sizes fail to account for the influence that information technology will have on their operations, business performance suffers (IT)[11]. Many businesses nowadays are creating or deploying their information technology systems (DSS), such as a decision support system, to maintain a competitive advantage in their respective industries (DSS) with respect to improving their Business Analysis and to better utilize Decision Support Systems at e-commerce in SMEs[12].

Few SMEs use information technology artifacts as strategic decision support (DSS) tools to better their business decisions. This has only recently become more widespread[13]. The absence of plans is most noticeable in how their websites have been implemented with modern technology's useful features[14]. The resources available to small enterprises (SMEs) are often somewhat restricted, and their managerial staff members have a lower level of expertise. Because of this, the likelihood of the collapse of a small business is increased because such companies do not always have access to the information they require[15], In specific knowledge about the operating environment of a business. Consequently, research on the design of IT artifacts for managerial decision support systems (DSS) is essential for improving business strategies[1].

In most cases, socioeconomic variables, rather than business changes, have been the focus of research on small enterprises (SMEs). There hasn't been a lot of research done on the requirements for decision-making (DSS) processes that micro and small firms have when developing IT products. Researchers tend to draw the same conclusions regarding experience and methodologies used in large organizations and apply them to small businesses, even though DSSs are extensively utilized. They do this without taking into consideration the fact that the DSS requirements for small enterprises are unique. This is still the case even though there are a great many distinct types of groupings[16]. Although strategic judgments are more crucial, most small businesses use decision support systems (DSS) to make tactical ones. Due to the unpredictability of their environments, small businesses have a hard time making long-business strategic decisions. Considering this when comprehending the intuition and judgment decision-makers employ in DSS design[13].

Many business processes can benefit from implementing E-Commerce because of the potential for increased efficiency and productivity. The challenges of modern E-commerce extend beyond technological ones into the realm of business. A decision support system (DSS) is "an interactive information system that gives information, models, and data manipulation tools to aid in making decisions in semi-structured and unstructured contexts." Online trade can take numerous shapes, each with its own set of advantages and disadvantages, costs, and levels of difficulty, depending on the needs of the business[15].

Multiple studies have shown that the e-commerce sector is highly competitive and constantly evolving. To thrive in today's market, firms must maintain a competitive edge through the introduction of new products and services, as well as the efficient management of operational processing using a decision support system. In addition, for businesses to maintain their success in the modern digital age, they need to adapt to the constantly advancing state of technology quickly[2].

Business Analysis and Decision Support Systems for e-commerce in SMEs are one of the most successful strategies, despite their inverted U-curve shape, fluctuating nature, and variety of approaches based on several factors. Business Analysis and Decision Support Systems in e-commerce in SMEs strategy could help businesses capture new ideas from external resources to keep up with the growing demand for their products[7].

If SME's access to the Decision Support System, E-commerce platforms can effectively manage their human resources, particularly those who work in IT support. The success of enterprises that operate as E-commerce platforms is mainly dependent on the IT support teams because the primary purpose of these teams is the collecting and process information[7].

Failures may be exceedingly costly in the information technology business, which places a significant amount of responsibility on the shoulders of the project manager to ensure that they do not occur. The authors of the study also stated that an agile project management system, provided with the appropriate level of oversight, can assist in ensuring that managers deal with problems as they emerge and refrain from providing incorrect updates on the status of their projects. The agile methodology places a strong emphasis on adaptability and rapid iteration when it comes to the process of developing a new and beneficial product or service. An IT project can be finished more quickly and with fewer complications, if its status is tracked correctly, according to [7], who found that this was also the case. This makes it easier for managers to avoid costly escalations and lowers the risk of failure. Therefore, scientific and methodical evaluations of product developments and IT project management in a dynamic environment are essential to improve the economics of doing business online.

If there will be a discussion about business analytics then simply it is stated as "the process of discovering meaningful and actionable insight in data." Business analytics is considered a relatively new term that is going to be famous among the nation to introduce various new and innovative ideas to enhance the knowledge of the individual for the accurate decision support system. If its history will be searched in the business world then there is nothing that can be searched and observed for this. Before starting a more specific discussion about business analytics and exploring its raking and stages for the process of decision support system, the analytics perspectives must be discussed that are much closer to the specific stage of decision making. Generally, business analytics is defined as the science and art of discovering, analyzing, and identifying novel insight from a wide range of data by considering sophisticated statistical models, machine learning, and mathematics for supporting more timely and accurate decision support systems at E-commerce [9].

Also, Things can benefit the e-commerce of SMEs by enhancing customer experience as well as providing reliable product and service delivery. Combining data can create new methods that can benefit SMEs. It is important to understand small and medium businesses contribute a large portion of a nation's gross domestic product. This indicates that these businesses have to maintain their growth. One way they can maintain this growth is through business analytics. The wide accessibility of business analytics has made it possible for small and medium businesses to adopt recent technologies to improve their performance and revenues. SMEs are making great steps by using the business analytics of decision support systems. With the use of smart technology, these businesses can efficiently utilize their resources, improve market share, as well as boost revenue. As such, many SMEs are using decision support systems in E-commerce as one of their business strategies.

Therefore, in the terms of business analytics, this discussion is all about decision support systems at E-commerce in SMEs as well as problem-solving. Due to the rapid growth of technology, business analytics can be stated in very simple a word that is "the particular process for discovering actionable and meaningful full insight in the data. The key objective of business analytics is to generate learning, knowledge, and understanding. Collectively, it is used to referee the particular insight for supporting the evidence that is based upon the decision-making for the process of performance management. By considering the aspects of evolution, it has been started several years ago and it also refers to the process of the competencies, practices, applications, and technologies that are being involved in order to attain such kinds of particular objectives. Therefore, several people are getting confused by merging the terms and concepts that are relatively similar to their meanings as well. Intentionally, sometimes it has been driven by vested interest and involved technology [10].

Today, in business decision-making is moving towards the particular point at which the accepted practices are all about the first understanding of the specific numbers as well as what they are revealing for it. It also includes the particular aspects that are utilized for this insight to deriving the decision of intelligent business. The approaches are also can be replaced where the action has been taken by people which is considered to feel better and right after that it is also used to identify the numbers that are used to analyze and see its work. Hence, the process of business analytics must be driven toward an accurate decision support system of E-commerce in SMEs. It also plays an essential role in the organizational landscape. This document explores the various aspects of business analytics as well as a decision support system in E-commerce. It also explores the positive effects of business analytics and decision support systems along with the negative aspects of business analytics and decision support systems. Both of these points and aspects are being discussed by considering the decision support system process [11].

## 2.1 Theoretical Content

There are many disadvantages of Business Analytics. There are many issues that can be described that can be mentioned for the main issue of why business analytics go wrong and such as when we make decision support system it can be wrong. There are many issues and problems that can be the main issue for that. We will discuss them one by one. One of the main issues is human judgment. Another main issue may be mentioned as a lack of governance of decision-making. If we do not mention another main issue like organizational culture then it will be a big problem for us. Now we will be in this position to describe what are the main and prominent issues and problems that can be the main resin of becoming business analytics become proof wrong. One of the main reasons is people.

## 2.2 People

People may be facing a different types of Bias on the basis of different levels of issues like age, gender, and ethnicity. Many people face, Bias. For example, a huge amount of people apply for the post of pacific jobs. But approximately a huge number of candidates have the same qualification, skills, and experience but they received a response from the organization to much more than the candidates of those people that applied with the name of African Americans names. So, we can now in this position to explain if the data set can be biased data set that it will consider it will be could be wrong. Because it is basically available on the basis of data set. That belongs to the biased data set [12].

## 2.3 Governance

Governance is also the main issue that can make Business Analytics become wrong. In governance, the main resign is Insufficient legislation. When a un insufficient legislation is available then it needs to be proper legislation for proper governance. Without law, it is very difficult to manage a single person, poor judgment, or an organization. Without proper legislation, it is impossible to manage of management of people. A proper way for the success of business analytics is to require governance and make in mind it is impossible to maintain governance. Social media need legislation. Now it is a big issue for humans. Without proper legislation, it is very difficult to manage individuals and manage accounts, and management and decision-making are impossible [13].

So, if proper governance and rules and regulation is available then it is possible that Business Analytics give a positive response otherwise it will give a negative result and all issue that tackle on the basis of Business Analytics become wrong. All over the world people spread propaganda against different religions on the basis of freedom. [14].

## 2.4 Culture

Culture is also the main issue for Business Analytics and decision support systems in E-commerce. If the culture of the organization will not support then different actions taken on Business Analytics will be negative. So, we need a proper and positive culture to maintain and implementation of proper Business Analytics. If the culture of the system. Now it describes the history that most organization required and tries to get maximization of profit. The organization tries to maximize profit. As a result of getting more profit quality of products or services is compromised. Day by day Quality is going to decrease and need. It is all about due to culture. So, we have to mature culture so that our different, Business Analytics will become positive [15].

## 2.5 Technology

Technology is also making a big impact on Business Analytics and decision support systems in E-commerce it is very important to take very careful to get data from systems or databases. Data on different social media can be created or hacked. after that, it can be used for a negative type of people so it is very important to secure data on any system. People have a very low level of human visibility. it is also a very low-level decision support system. it also depends on a different ethical basis.

There is an important link between the use of the human decision support system and business analytics. This link has been neglected oftentimes despite such importance. The research on the connection between business analytics and decision support systems.

Positive arguments for business analytics and decision support Systems

In order to explain and explore the benefits and positivity of business analytics and decision support systems in E-commerce three important points are discussed. These three aspects are people, governance, and culture. The aspects of business analytics and decision support systems also require introducing positivity to the requirements. It has been analyzed that business analytics must be coordinated across the groups of several stakeholders to empower the trust gap which is existed currently among the parties. In order to do this as well as perform this in a good way, humanity and society will be prepared for responding to the business analytics that is related to the ethical demands and that will have required to grow variety and complexity in the future.

## 2.6 People

If the aspects and points of the decision-support system and business analytics in E-commerce must have been discussed to highlight the individuals who are involved in this particular process of business analytics. First of all, the discussion is entirely about the professionals of business analytics who are accountable for this morally as well. It has been indicated in various studies that the technology of big data is neutral ethically. Even sometimes, it is not considered a built-in perspective to explain what is right or wrong and what is bad or good in its utilization. There are several codes of ethics that existed relevant to the computer sciences and data analytics. By adopting all of these data analytics businesses can get access in good ways. Even if the individual will be utilized in their organization the business performance can be easier as compared to various other aspects. For the profession of analytics, the absence of the framework can be the cause of leaving voiding by which individuals are not clear on the particular consistent standards for their productivity within the organization and working with their colleagues as compared to other disciplines or industries [16].

## 2.7 Governance

This section is entirely a discussion about the rules, laws, and regulations that can be enforced for business analytics in decision-making. It has been argued by the houses of cultures, common digital, support committees, and media that for the technology companies the era of self-regulation must end. The SMEs were trusted to self-regulate for too long a process in the field of growing data and analytics SMEs. In the field, the public interest has been undermined as well as highlighting the several benefits of analytics. It also offers an exclusive overlook to society; it includes exceptional user experience as well as fraud prevention. One of the most important positive steps is the implementation of the GDPR general data protection regulation for protecting the individual's privacy rights in the EU. A powerful message has been sent out by regulation for the process of the data related to the level of care that is expected at the time of the handling of data of individuals

## 2.8 Overall culture

The culture is discussing the documented ethical principles that are required for enhancing the productivity of the organization. Additionally, ethically minded business analytics can be employed through various methodologies. The culture of careful ESR can be reinforced in the SMEs.by considering the publication and development of ethical principles that have been utilized as the points of the references that are against the framing of the analytics activities.

Summing up all the discussion it is concluded that business Analytics plays a very important role in the progress of the systems. But it will become negative if it is implemented without the proper way. It became wrong and gives negative results if it is not managed properly. We discuss in detail if people have to face a different level of baseness like gender, discrimination on skin colour, or if the business is based on belonging to a different culture and gender then will going go wrong for diction making and business analytics become a not good decision support system way [17]. The governess is also taken an important role in positive decision-making way. In governess, it also needs proper legislation and proper way of Rules and regulations. It is not possible without proper rules and laws required for social media for the respect of other religions.

## 3. LIMITATIONS AND FUTURE

The present work aimed to investigate the impact of business analytics and decision support systems on e-commerce in SMEs. Business analytics are highly adopted in business organizations to improve decision-making systems. Business analytics are useful for e-commerce businesses as they can provide information about online shopping trends and comprehend shifts in customer behavior. Data-driven decisions are usually beneficial for business organizations to help them improve their user experiences and expand business operations in profitable directions. Although, the appropriate selection of these analytics and support systems is possible through detailed research projects. Therefore, a research project studying business analytics and decision support system is highly important for business organizations dealing with online shopping or e-commerce business operations. In this research, the main emphasis is given to secondary information collection using authentic and relevant research articles and other sources from academic literature. Even the several measures were taken into consideration to determine the high authenticity, originality, fairness, and accuracy of the research outcomes still this research project is lacking in some areas. Lacking is representing critical issues that can influence the research outcomes or at least comply with the ethical standards set for a high-quality research paper in the respective field. In this section, these issues are discussed as limitations of this research project. Followed by these limitations, this section will also discuss future work.

Starting with the research project limitations, there are three main limitations of this research project which are discussed here. Firstly, the entire focus of this research project is on two variables only. Sometimes, there can be other mediating factors and variables which may have influenced the outcomes and results for these selected variables of research interest (business analytics and decision support systems). Regarding business analytics, its suitability and adaptability for an e-commerce business are essential for meaningful research findings. Otherwise, inappropriately selected business analytical approaches and methodologies can inappropriately influence their impact on e-commerce business operations executed by small and medium-sized business organizations. In this research project, we did not give sufficient attention to various business analytics and relevant suitability for e-commerce businesses. The obtained information is presenting a general scenario only. In the selection process, we only checked whether the research project contains knowledge of business analytics in e-commerce (SME) businesses or not. We did not define this criterion in more detail to study other factors and variables such as adaptability and suitability which represents a major limitation of this project [18].

The second major limitation associated with this research project is linked to the data collection methodology and research designs. In this project, we have obtained secondary information from the literature to extract research findings on this topic. Relying on secondary research data is not sufficient to originate actual research findings. Regarding such important research topics, a mixed-method approach could be utilized to enhance the relevancy, validity, and reliability of the research findings. In the research project, secondary research data should be combined with primary research to critically assess the actual impact of business analytics and decision support systems in e-commerce businesses in the SME sector. Primary research data could be extracted by using the interview or survey approaches. Regarding this relevant e-commerce companies could be invited to share their experiences with business analytics and decision support systems (if they have implemented both or at least one of these variables at their workplace). A combination of secondary and primary research data can cover this research limitation [18].

The third and most important limitation of this research project is linked with the generalization of extracted research findings. In this research project, we have not classified e-commerce companies into different geographical segments or countries.

Variation in geographical locations relates to the economic situation and variation in social factors (e.g., familiarity with advanced technology, the tendency towards technological adoption, and awareness levels). Diversity in geographical locations emphasizes the inclusion of geographical details in the research project. Obtaining information from various research articles may not satisfy the requirements for research findings generalization. The inclusion of research articles from diverse geographical segments may have the tendency to a specific geographical segment or county if articles included from that segment or country are greater in number. In this situation, the generalization of research findings will not meet the high reliability and validity status. Thus, it represents the major limitation of the research project [19].

Moving to future work, a new research trend is evident in existing literature under which researchers are obtaining primary research findings from original respondents regarding their challenges and adopted mitigation strategies. In these research studies, researchers are emphasizing the identification of various challenges experienced by the management of e-commerce businesses in the SME sector. These challenges will be related to the adoption, selection, and assessment of business analytics and decision support systems at the workplace. In the future, we can also direct this research project in this direction. Somehow, we can also include new variables of research interests in future work to promote its usability in academic and commercial sectors. Research findings of the future research project will aware the management of e-commerce businesses take effective decisions regarding the selection of business analytical approaches and systems while understanding the expected challenges behind this huge change. Moreover, it will open a new path for future researchers to study challenges and issues with strategic solutions in the SME sector while studying the cases of e-commerce businesses dealing with business analytics and decision support systems. In the future, we will also work on a separate research project indicating the situation of the same research variables with a specific geographical context. We aim to study e-commerce SME businesses adopting business analytics and decision support systems in the Middle East region. Currently, the middle east is an emerging market with high potential for e-commerce business. Several new e-commerce businesses from small and medium-sized sectors have recently expanded their businesses as large corporations with B2B business models. This situation is an indicator of high potential in this market segment. Consequently, considering the growing market of e-commerce business in Middle East emerging markets we will focus our future research project on this segment. Conclusively, a research project on the Middle East will add worth to the existing literature and bring forth applicable findings for the commercial sector (e.g., management of e-commerce businesses from the SME sector) to grow their businesses and sustain their growth trends.

## 4 RECOMMENDATION

Considering the above discussion, literature review findings, limitations, and future work a few recommendations are stated below. These recommendations are subdivided into two parts. In the first part, recommendations are presented for the audience or readers of this research project. On the other hand, recommendations shared in the second part of this research project are for researchers to improve their research work for the future.

### *4.1 Recommendations for the Audience*

Firstly, eCommerce management should try to record their sales transaction and frequency of sales data in internal databases. This data can be supportive of the data analeptics. In the business analytics of e-commerce businesses, managers are supposed to use historical data for the identification of patterns and trends to discover new findings related to their executed business operations and customer buying patterns. Storing data in secure internal databases with historical series can be beneficial for the management of e-commerce businesses to use in business analytics. Without having sufficient data, business managers may not be able to extract meaningful and highly reliable information for their business. For example, if the offered product is linked with seasonality then data from one year is not sufficient. Managers should make sure to keep records of sales data and other informative content for more than one year to identify each trend in seasonality. Thus, using business analytics managers of e-commerce businesses will be able to estimate future demands and requirements for inventory or stocking. This kind of information is also useful in determining the reorder point and improving the supply chain decisions to smoothly complete all business operations.

Secondly, data regarding electronic commerce should be also converted into categorical variables for better visualizations of results in the decision support system. For example, if an e-commerce business is offering two flavors for a single product, then both should be presented in the same graph or chart to determine customer preferences. Such visualizations (e.g., pivot tables, pivot charts, and time series charts) are useful in the decision-making process for the future direction and modifications of the e-commerce business.

Thirdly, academic researchers working on a similar research project should consider earlier discussed research limitations. The information available in this research article is generalized using secondary research data which may have differences from the actual research findings of the selected geographical context of the next researchers. Therefore, researchers are advised to consider the above limitation of this research project to enhance the chances of producing a highly reliable and valid research project for the academic literature. Furthermore, some findings may also get change with the passage of time. Currently, the technological situation is dynamic therefore, we are uncertain about the future impact of this research project. Future researchers using this project for their research project should also consider this issue. Research findings would be useful for general information only therefore, these findings should not be based on a research project in a different context while ignoring the social and technological variabilities.

Fourthly, research findings suggest that e-commerce business managers should also cover diversity-based methods if business operations are linked with infrequent purchases. Seasonality and infrequent purchases from these e-commerce websites sometimes cause challenging situations for business managers to extract meaningful findings from the qualitative data to further use in the decision-making process. In this situation, business managers should also consider relevant algorithms and business analytics approaches that comply with infrequent and seasonal sales.

#### *4.1.1* **Recommendations for Researchers**

This section is entailing information about key recommendations for the researchers to improve their research projects in the future. These recommendations are mainly drawn by considering the limitation section as these limitations are required to be eliminated and properly managed for a future research project. The most important and relevant recommendations for the researchers are stated below:

Firstly, researchers should further research this topic by continuously studying variations in business analytics and decision support systems in the same context. Researchers should cover this information by using the empirical research approach. A number of e-commerce websites should be selected by using clearly defined research criteria to collect original responses from the research population. In the defined criteria researchers should cover e-commerce business size, business model, and suitability for the advanced decision-making processes. The selected businesses should have experience with at least one of these: business analytics or decision support systems. In the selection process companies having long-term experience should be preferred to companies having short-term experience with business analytics and decision support systems. An appropriate selection of e-commerce businesses for the SME sector will be beneficial for this research project in the future.

Secondly, researchers should also include information about the selected research articles in a systematic literature review format. Under this format, researchers should share information about the research groups, county, research journal, and year of publication. Such information is highly important in a secondary research project.

Thirdly, future research projects should also include information about other research variables which can work as mediators for this research project. Regarding decision support systems, researchers should also cover details on security measures and management systems at workplaces for the obtained data from e-commerce websites. Moreover, a detailed research project in the future should also cover fuzzy decision support systems for the risk assessment and management of e-commerce businesses in the SME sector. Real-coded genetic algorithms used in e-commerce businesses are also supportive of the implementation and functionality of fuzzy neural networks which need to be further studied while determining the impact of decision support systems and business analytics on e-commerce businesses [20]. Thus, by working on these recommendations this project can be made a better and improved version for future researchers and research audiences interested in this topic.

## 5. CONCLUSION

The widespread adoption of E-Commerce has the potential to boost the productivity of a great many business activities. The technological and commercial challenges of e-commerce are significant in businesses globally. Solving one of these issues is not easy. The term "decision support system" refers to "an interactive information system that delivers the information, models, and data manipulation capabilities to assist in making decisions in semi-structured and unstructured settings." This is so since "an interactive information system that delivers the information, models, and data manipulation capabilities to aid in making decisions in semi-structured and unstructured settings" Decision support systems are defined as "information systems that give information, models, and data manipulation tools to aid in the process of making decisions" . There are a variety of approaches to e-commerce that can be used depending on the requirements of the business; each may involve a varying degree of complexity and cost. Depending on the method selected, the business may incur the quantity of either.

To make informed decisions regarding the day-to-day operations of their firms and the long-term business plan for those organizations, decision-makers in small businesses need to have access to effective decision support systems. These decisions could mean the difference between business and failure for the companies. The decision-making technologies that are currently available give a wide variety of system provisions for managerial decisions. They continue to have a limited capacity to cope with the cognitive business requirements of decision-makers, which is particularly problematic for small businesses competing in the online consumer market. Today's small enterprises must still contend with various obstacles to survive and thrive in their respective industries. Some of these difficulties are attributable to the ever-changing nature of the business climate, the continually rising capabilities of technology, and the ever-evolving tastes of clients. Small business needs to have an interactive website that gives them access to a platform comparable to what larger organizations have for them to tackle these problems. Often, the owners and managers of small businesses may not have access to the information required to maintain their current websites. This can make it difficult for them to keep their websites up to date. Because of this, it may be challenging for them to maintain the information on their websites current. More than half of the retail establishments take part in promotional activities on the websites of local small companies. This percentage is higher than it is in the United States. Even though it is essential for managers to have a digital strategy that uses effective websites, social media, and mobile applications, most owners and managers do not have confidence in their ability to make strategic judgments regarding fundamental problems such as this one. This is even though managers need to have a digital strategy.

Many managers still do not have a digital strategy, even though this is exceptionally vital for them to have. The impact of Business Analysis and Decision Support Systems in e-commerce in SMEs is one of the best strategies for growth and development. In relation to designing an exclusive managerial support system in the operation of e-commerce for SMEs, a vital task would be to conduct a design study in future for capturing insights of the decision-making problems [for example, 21, 22], through adopting modern research methodology, such as design science principles [23, 24, 25, 26, 27].